\documentclass[a4paper]{jpconf}
\usepackage{graphicx}
\begin{document}

\title{Large scale simulation of quantum-mechanical molecular dynamics  
for nano-polycrystalline diamond}

\author{Takeo Hoshi$^{1,2,3}$, Toshiaki Iitaka$^{3}$ and Maria Fyta$^{4}$}

\address{(1) 
Department of Applied Mathematics and Physics, 
Tottori University, 4-101 Koyama-Minami, Tottori 680-8552, Japan}
\address{(2) Core Research for Evolutional Science and Technology, 
Japan Science and Technology Agency (CREST-JST), Japan}
\address{(3) RIKEN (The Institute of Physical and Chemical Research), 2-1, 
Hirosawa, Wako, Saitama 351-0198, Japan.}
\address{(4) 
Physics Department, 
Technical University of Munich, James Franck Strasse, 85748 Garching, Germany.}

\ead{hoshi@damp.tottori-u.ac.jp}

\begin{abstract}
Quantum-mechanical molecular-dynamics simulations 
are carried out to explore possible precursor states
of nano-polycrystalline diamond, 
a novel ultra-hard material produced directly from graphite. 
Large-scale simulation with 10$^5$ atoms is realized
by using the \lq order-$N$' simulation code \lq ELSES' (http://www.elses.jp).
The  simulation starts with a diamond structure that contains initial structural defects
and results in a mixture of graphite($sp^2$)-like and diamond($sp^3$)-like  regions as nano-meter-scale domains.
We speculate that the domains are metastable and 
are possible candidates of the precursor structures.
\end{abstract}

\section{Introduction}

Nano-polycrystalline diamond, 
a novel ultra-hard material, 
is produced directly from graphite at high temperature and high pressure and
is of great interest both for fundamental science and industrial applications.
\cite{IRIFUNE-NPD-NATURE2003, SUMIYA-NPD-HPR2006}
Nano-polycrystalline diamond
consists of fine diamond crystals of 10-30 nm in size
and has a characteristic lamellar structure in the 10 nm scale. 
Its precursor structure is crucial for the controllability of 
the growth process and the functions of nano-polycrystalline diamond and 
was investigated, for example, in a recent experiment. \cite{GUILLOU-2007} 

In the present paper, nano-polycrystalline diamond is explored 
by large-scale simulations based on quantum mechanical molecular dynamics 
\cite{ELSES-URL, TAKAYAMA2004, TAKAYAMA2006, HOSHI2006,HOSHI-JPCM2009}, 
with up to $10^5$ atoms. 
The aim is to find candidates of intrinsic precursor structures of nano-polycrystalline diamond, 
which cannot be achieved in smaller scale simulations.
The present work is the first stage for simulating a controllable growth process,
since a simulation of growth process requires proper initial structures 
and intrinsic precursors will give such initial structures. 

\begin{figure}
\begin{center}
\includegraphics[width=0.85\linewidth]{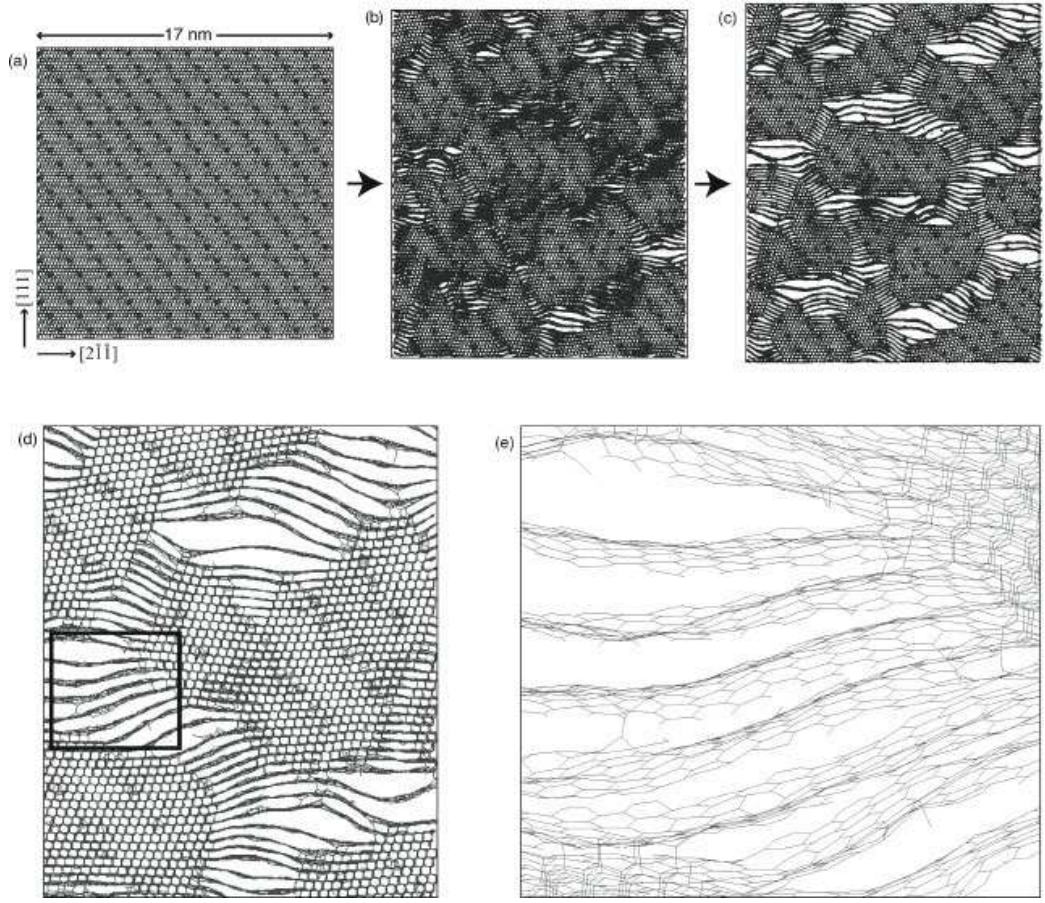}
\end{center}
\caption{
\label{fig-1}
(a)-(c): Successive snapshots of a simulation 
with 107,520 atoms under the [111] tension, at time steps 0, 720 and 820.
(d): Close up of the upper right region of (c).
(e): Close up of a graphite-like region shown by the rectangular box in (d). 
 A diamond-like region appears in the upper right area.
The view point in (e) is slightly tilted from the one in (a)-(c). 
}
\end{figure}

\section{Method}

The large-scale simulations were realized by an \lq order-$N$' 
electronic structure calculation based on Krylov subspace theory,
in which the computational time scales linearly with the system size.  
\cite{TAKAYAMA2004, TAKAYAMA2006, HOSHI2006}
In the Krylov subspace theory,
the  Green's function  is calculated, instead of eigenstates. 
The methodology has a rigorous mathematical foundation
as iterative linear-algebraic algorithms and 
is applicable to both insulators and metals.

In general, an \lq order-$N$' calculation is represented in real-space.
The one-body density matrix is defined as
\begin{eqnarray}
\rho(\vec{r}, \vec{r} \,') \equiv \sum_i f_i \phi_i^{\ast}(\vec{r}) \phi_i(\vec{r} \, '),
\end{eqnarray}
with eigenstates $\{ \phi_i \}$ and occupation numbers $f_i$, 
where $\vec{r}, \vec{r} \,'$ are the positions.
Then the expectation value of 
a physical quantity $\langle X \rangle$ is calculated by
\begin{eqnarray}
\langle X \rangle = 
{\rm Tr}[\rho X] = \int \int  \rho(\vec{r} \, ', \vec{r} \,)^{\ast} 
X(\vec{r}, \vec{r} ')   \, d \vec{r} \, d \vec{r} '. 
\end{eqnarray}
In the Krylov subspace theory,
the Green's function $G=G(\vec{r}, \vec{r} \, '; \varepsilon)$
is calculated by an iterative algorithm 
and the density matrix is given by
\begin{eqnarray}
\rho(\vec{r}, \vec{r} \, ') = - \frac{1}{\pi} \, \int \, f(\varepsilon) \,
{\rm Im} \left[ G(\vec{r}, \vec{r} \, ';\varepsilon+ i 0) \right] \, d \varepsilon 
\end{eqnarray}
where $\varepsilon$ is the energy
and $f(\varepsilon)$ is the occupation number given by the Fermi-Dirac function.
Now the simulation code is being reorganized 
as a package with the name \lq ELSES'
(=Extra-Large-Scale Electronic Structure calculations). \cite{ELSES-URL}
The complete reference list of ELSES for its methodology and application
is found in a recent paper. \cite{HOSHI-JPCM2009}
In this paper, 
a tight-binding form Hamiltonian for carbon \cite{XU} is used.
The methodological details of the Krylov subspace theory 
in the present simulation 
are the same as those for liquid carbon. \cite{HOSHI2006}

In the first stage of our research, the simulations were carried out 
by expanding a diamond structure with initial structural defects, 
a small fraction of threefold-coordinated atoms 
and deformed fourfold-coordinated atoms, not more than 10 \%.
In this way, we aim to obtain the intrinsic metastable structures of 
nano domains within the limit of the computational time scales. 
Three-dimensional periodic simulation cells are used for the samples.
The samples were prepared as super cells of smaller samples 
with initially structural defects.
\cite{FYTA-PRL-2006}
The procedure of generating the initial structure is given in a paper 
\cite{FYTA-PRB-2003}
The computational time with 107,520 atoms is typically 
 20-23 minutes per molecular-dynamics step on a standard workstation that has two quad-core 
 Intel Xeon$^{\rm (TM)}$ processors (E5345, 2.33GHz).

\section{Results and discussion}

Figure ~\ref{fig-1} shows 
a simulation result with 107,520 atoms under the [111] tension. 
The snapshots of 
Fig.~\ref{fig-1}[(a),(b),(c)] correspond to 
the  initial, 720-th and 820-th steps, respectively.
The time interval per molecular-dynamics step is $\Delta t = 3$ fs and
the last step corresponds to the elapse time of approximately 2.5 ps.
An orthorhombic cell is used as the periodic simulation cell
and  the initial cell lengths are 17.4, 17.6 and 2.0 nm
along the $[111]$, $[2 \bar{1} \bar{1}]$ and $[01\bar{1}]$ directions, respectively. 
Fig.~\ref{fig-1} depicts the atoms in the simulation cell
with a viewpoint along the $[01\bar{1}]$ direction.
The [111] tension is realized by expanding 
the cell size only for the $[111]$ direction.
The cell length is expanded by $0.03 L_0$ every 100 steps,
where $L_0 = 17.4$ nm is the initial length. The structure relaxation for given cell lengths 
was carried out under thermal fluctuations, introduced by a Nos\'e thermostat \cite{NOSE}.
The temperature of the thermostat is set to be $T=600$K.

The simulation resulted in a mixture of graphite-like  regions 
in the $sp^2$ state and diamond-like regions in the $sp^3$ state, 
as in Fig.~\ref{fig-1}(c). 
Several initial defects introduce graphite-like regions,
since defective regions are relatively unstable and 
easier to be transformed into other stable or metastable forms.
The graphite-like sheets are perpendicular to the [111] direction and are "wavy"
and the diamond-like domains have several characteristic domain shapes.
Similar domain shapes are formed in simulations 
with smaller numbers of atoms ($10^3-10^4$atoms) and a longer time scale ($10^1$ps). 
We speculate that these domain structures are metastable 
and are candidates of possible nano-polycrystalline domains or 
their precursor structures. 

\begin{figure}
\begin{center}
\includegraphics[width=0.6\linewidth]{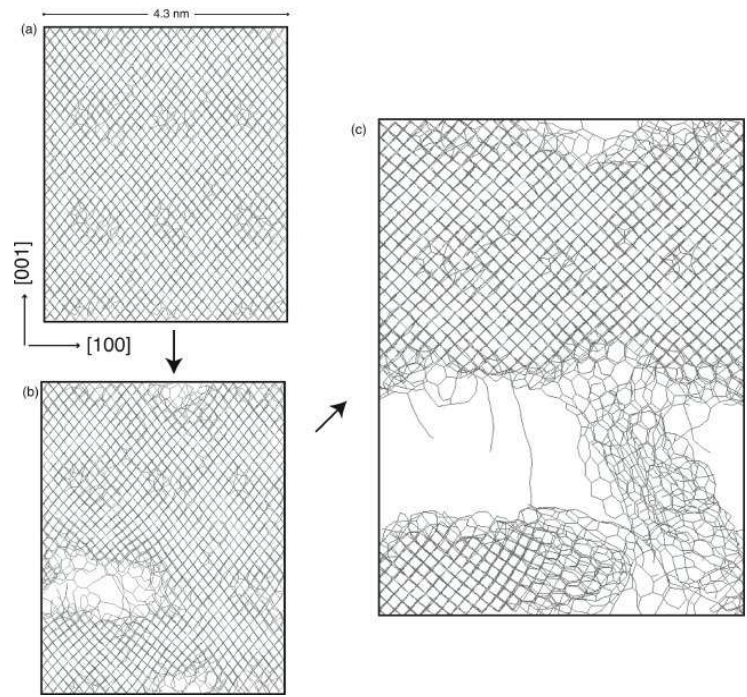}
\end{center}
\caption{
\label{fig-2}
(a)-(c): Successive snapshots of a simulation with 4,608 atoms under the [001] tension,
at time steps 2,500, 2,900 and 3,500.
The last snapshot (c) is magnified by 50 \% in its size to clarify the structure.
}
\end{figure}

Simulations were carried out also under the [001] tension to confirm that 
graphite-like regions appear not only under the [111] tension.
Figure ~\ref{fig-2} shows a simulation with 4,608 atoms under the [001] tension.
An orthorhombic cell is used as the periodic simulation cell
and  the initial cell lengths are 4.3, 4.3 and 1.4 nm
along the $[100]$, $[010]$ and $[001]$ directions, respectively. 
The time interval per molecular-dynamics step is $\Delta t = 3$ fs.  
The [001] tension is realized by expanding the cell size only along the $[001]$ direction.
The cell length is expanded by $0.01 L_0$ every 100 steps,
where $L_0 = 4.3$ nm is the initial length. Fig.~\ref{fig-2} [(a),(b),(c)] correspond to 
the 2500-th, 2,900-th and 3,500-th steps, respectively. In the last snapshot, Fig.~\ref{fig-2}(c),
several graphite-like six-member rings appear at the right lower region.

In summary, we have shown that 
the order-$N$ electronic structure theory realizes
large-scale calculations with $10^5$ atoms on a standard workstation and
can play a crucial role in the investigation of nano-polycrystalline diamond.
In the future, a more systematic investigation, with quantitative analysis of atomic and electronic structures, will be carried out to serve as a direct comparison with experiments.

\section*{Acknowledgements}

This research was supported partially by Grant-in-Aid for Scientific Research on 
Innovative Areas 
\lq Earth Science Based on the High Pressure and Temperature Neutron Experiments' 
(No. 20103001-20103005), 
from the Ministry of Education, Culture, Sports, Science and Technology (MEXT) of Japan.
Numerical calculation was partly carried out using the supercomputer facilities of 
the Institute for Solid State Physics, University of Tokyo and the Research Center for 
Computational Science, Okazaki.


\end{document}